\documentclass[prd,aps,tightline,nofootinbib,showpacs,superscriptaddress]{revtex4}
\usepackage{graphicx}
\usepackage{amssymb}
\usepackage{amsmath,bm}
\usepackage{afterpage}
\usepackage{float}
\usepackage{epsfig}
\usepackage{subfigure}
\usepackage{ulem}
\begin{document}

\title{RGE, the naturalness problem and the understanding of the Higgs mass term}

\author{Ligong Bian}

\email{lgb@mail.nankai.edu.cn}

\affiliation{School of Physics, Nankai University, Tianjin 300071, P. R. China}

\date{\today}

\begin{abstract}

The naturalness
problem might be studied on the complex two dimensional plane
with the technique of dimensional regularization(DREG).
The Renormalization group equation(RGE) of the Higgs mass on the plane suggests the Higgs mass
approaches zero at ultraviolate (UV) scale, the scale can be Planck scale
when the top quark pole mass $M_{t}=168$ GeV.
The real issue of the naturalness problem in the sense of
Wilsonian renormalization group method is not about quadratic divergences
but the rescaling effect.
The Higgs mass can be considered to be one composed mass. All terms in the
lagrangian in this scenario are marginal terms and no relevant terms are left,
thus no rescaling effect
to cause the naturalness problem anymore.
RGE of the vacuum expectation value (VEV) in the Landau gauge
up to two-loop order is studied. Scale-dependent behavior of the composed
Higgs mass shows that we can have one tiny Higgs mass at high energy scale,
even around the Planck scale,
when $M_{t}\leq170.7$ GeV.

\end{abstract}


\maketitle

\section{Introduction}

In the quantum field theory, the naturalness  problem assumes new physics around
TeV scale. While, no new physics sign has been observed at LHC and no indications have been found
which indicates the SM could be a low energy effective theory below the Planck scale sofar.
Meanwhile, recent studies on the stability and (meta)stability,
with the current Higgs-like mass observed at LHC, suggest that the SM may apply to
the Planck scale~\cite{Alekhin:2012py}.
Confront this situation, the naturalness problem need to be revisited.
To investigate the naturalness problem, two important issues are the way to understand
the Higgs mass term and quadratic divergences.

Firstly, we consider the case that the Higgs mass is one gauge invariant quantity.
To reveal quadratic divergences, one straightforward way is to use cut-off method,
when we calculate quantum corrections to the mass term.
The cut-off introduced can be the cut-off in a UV complete theory or the energy scale
at which new physics enter into the low energy physics.
 Quadratic divergences can also been manifested with the DREG method.
At one-loop order, a suitable criterion to consider quadratic divergences properly
within the framework of DREG, is the occurrence of poles on the complex
two dimensional plane~\cite{Veltman:1980mj}.
We investigate quadratic divergences calculation with cut-off method,
and the way to reveal quadratic divergences with DREG on the complex $d$
dimensional plane at any loop orders. The connections of these two methods are also explored.
The quadratic divergences of one-loop level dominates quadratic divergences
of the quantum corrections to the Higgs mass term.
And with quadratic divergences of one-loop order, the naturalness problem
was studied in 't Hooft-Feynman gauge completely three decades before~\cite{Veltman:1980mj}.
In this work, we explore if the problem is gauge dependent through proceeding calculations
in $R_\xi$ gauge.
The naturalness problem prevents us from calculating the high energy behaviors of the SM
up to UV scale. We find that with physics relating to quadratic divergences being attributed
to the complex two dimensional plane, one can safely study renormalization group (RG)
behaviors of physical parameters with DREG and $\rm{MS}$(or $\overline{\rm{MS}}$) scheme.
After we develop the method to express physical parameters on the complex two dimensional plane
through parameters of the $d=4$ case, we derive the RGE of the Higgs mass on the plane,
wherein the Veltman condition can be expressed by running parameters consistently
in our renormalization procedure.
With energy scale increasing, the Veltman condition is found to be fulfilled at UV scale, and the
the scale at which the Higgs mass on the complex two dimensional plane approaches zero
is slightly higher.
Ward identities do hold for all the complex dimensions within
DREG~\cite{'tHooft:1971fh,'tHooft:1972fi},
thus our study in this part is manifestly gauge invariant.

In the
Wilsonian renormalization group method, quadratic divergences is not the real issue
of the naturalness problem and can be absorbed to fix the position of critical surface~\cite{Aoki:2012xs}.
The role played by the surface is very resemble with that of complex $d$ dimensional plane.
For the possibility that the Higgs mass is considered not to be an gauge invariant but one composed quantity,
the rescaling effect which causes the naturalness problem
in the sense of Wilsonian renormalization group method is absent, and then we need not
to worry about the naturalness problem. The RGEs of VEV and
the Higgs term in this situation will be studied.

This paper is organized as follows. In section~\ref{sec:dim},
we study the relation between complex $d$ dimensional plane
and quadratic divergences, then we give one method to express the physics on the
complex two dimensional plane with parameters of $d=4$ case in DREG.
The naturalness problem in the context of the SM is derived in $R_\xi$ gauge.
With DREG to proceed divergences and using $\rm{MS}$ (or $\overline{\rm{MS}}$)
scheme as the renormalization scheme, the RGE of the Higgs mass on the
complex two dimensional plane
at one-loop level in the sense of~\cite{Veltman:1980mj} is derived,
and been expressed by parameters of $d=4$ case in section~\ref{sec:rg2}.
In order to make our results general, all calculations in this work are
proceeded in $R_{\xi}$ gauge.
In section~\ref{sec:wilRG}, we revisit the meaning of divergences
in Wilsonian sense~\cite{Wilson:1973jj,Peskin:1995ev}.
With the help of the Higgs mechanism, the theory with only marginal
terms left is considered, in which the
mass term is replaced by the composed mass. And after we generalize this
mind to the SM, the RGE of VEV is derived. Then, we take one numerical
analysis of the RGE of the composed Higgs mass together with beta functions
up to two-loop order.
These construct section~\ref{sec:cmH}.
Our conclusions are given in section~\ref{se:conclusion}.

\section{Complex $d$ dimensional plane and quadratic divergences}
\label{sec:dim}

The concept of analytic continuation in the
number of dimensions is the basis
of the DREG method~\cite{'tHooft:1972fi},
where the dimension $4$ is generalized to be
the complex $d$,
and the ultraviolet infinities in the original four dimensional
momentum integrals manifest themselves as poles on the complex
$d$ dimensional plane. As for the mathematic meanings of analytic continuation
and complex $d$ dimensional plane, one can refer
to~\cite{Leibbrandt:1975dj}. To find physical meanings of the
complex $d$ dimensional plane, especially the complex two
dimensional plane, we investigate the quantum corrections to
the mass term with cut-off and DREG method together.

\subsection{Quantum corrections to mass term: quadratic divergences and complex $d$ dimensional plane}

A direct relation between
power counting and the location of the poles on the complex
$d$ plane can be found with the analytic continuation method.
Quadratic divergences and logarithmic divergences
correspond to poles at
$d=4-2/L$ and $d=4$, with $L$ denotes the number of loops ~\cite{'tHooft:1973pz}.
In following paragraphs,
we illustrate this relation by investigating one- and two-loop
order situations.

At first, we study the one-loop situation which is useful to derive
the formula to express naturalness problem in section~\ref{sec:rg}.
In the ordinary perturbation calculations of $\lambda\phi^{4}$ theory,
quantum corrections to mass term can be figured out
from momentum integral calculations with cut-off method.
When the mass term is not
negligible but much less than the fundamental scale $\Lambda$ and the
lower energy scale
$\mu$\footnote{This $\mu$ should be different from the scale parameter $\mu$ introduced
in $\rm{MS}$ scheme when we derive RGEs of physical parameters.},
i.e., $\Lambda\gg\mu\gg m$, the momentum integral is divergent as
\begin{eqnarray}\label{eq:cutoff}
\lambda\int\frac{d^{d}k}{(2\pi)^{d}}\frac{1}{k^{2}-m^{2}}
\rightarrow\bigg\{\begin{array}{c}
\frac{-i\lambda}{4\pi}\log\frac{\Lambda^{2}}{\mu^{2}}\;,
\qquad\qquad\qquad\qquad  d=2\;, \nonumber\\
\frac{-i\lambda}{16\pi^2}\big(\Lambda^{2}-\mu^{2}-m^{2}\log
\frac{\Lambda^{2}}{\mu^{2}}\big)\;,\qquad d=4\;.
\end{array}
\end{eqnarray}
With which
and the renormalization prescription as will be explored in section~\ref{sec:rg2},
the relationship between
bare and renormalized mass in the four dimensional $\lambda\phi^{4}$
theory can be written as
\begin{eqnarray}\label{eq:massphi4}
m^{2}-\frac{\lambda}{16\pi^2}\big(\Lambda^{2}-\mu^{2}-m^{2}
\log\frac{\Lambda^{2}}{\mu^{2}}\big)
=Z_{\phi}m^{2}_{0}\;.
\end{eqnarray}
The quantum contribution $\Lambda^{2}-\mu^{2}$ constructs the source
of the naturalness problem in
our four dimensional field theory, as will also be shown in Eq.~(\ref{eq:DRED}).

Poles of momentum integrals, which are involved in
self-energy computations and calculated with DREG in $d$ dimensional spacetime, are shown below and listed in Appendix~\ref{sec:IDC},
\begin{eqnarray}\label{eq:dreg}
\lambda\mu^{\varepsilon}\int\frac{d^{d}k}{(2\pi)^{d}}
\frac{1}{k^{2}-m^{2}}
\rightarrow\bigg\{\begin{array}{c} \frac{i\lambda\mu^2}{4\pi(d-2)}\;,\qquad d\rightarrow 2\;, \\
\frac{-i\lambda m^2}{16\pi^2(d-4)}\;,\qquad d\rightarrow 4\;.
\end{array}
\end{eqnarray}
Where the arbitrary scale parameter $\mu$
introduced is to give the running scale of
RGEs in $\rm{MS}$ (or $\overline{\rm{MS}}$) scheme.
$\mu^{\varepsilon}$ with $\varepsilon=4-d$ compensates the dimension of $\lambda$
in $d$ dimensional Lagrangian. And from the above equation,
one can achieve poles at $d=2$ or $d=4$ when the dimension $d$ continues to $4$ or compacts to $2$. And these two kinds of poles could not emerge simultaneously.

Compare Eq.~(\ref{eq:dreg}) with the Eq.~(\ref{eq:cutoff}), the
correspondences
\begin{eqnarray}\label{eq:correspondence}
\frac{1}{1-d/2}\rightarrow\frac{\Lambda^{2}}
{4\pi}\;,\quad
\frac{1}{2-d/2}\rightarrow\ln\frac{\Lambda^{2}}{\mu^{2}}
\end{eqnarray}
can be obtained, the quadratic and logarithmic divergences
correspond to poles at $d=2$ and $d=4$ on the
complex $d$ dimensional plane.
When the dimension $d$ continues to $4$, logarithmic physics can be given.
And quadratic divergence might be considered as the physics on the complex two dimensional plane,
which will also be explored in the next subsection.

With DREG adopted as
the regularization method, the quadratic divergences manifest as the pole at $d=2$, i.e., the pole on the so-called
complex two dimensional plane.
The quadratic divergences part of quantum corrections to the Higgs mass
at one-loop order can be calculated with this method in the context of the SM.
With which, the formula to express the naturalness problem in 't Hooft-Feynman gauge is given in the paper~\cite{Veltman:1980mj}. And we derive the formula
in $R_\xi$ gauge in this work.

Now we move onto the two-loop level situation.
Momentum integrals, which give quadratic divergences part
of quantum corrections of the mass term in $\lambda\phi^4$ theory and the SM,
take the forms of
\begin{eqnarray}\label{eq:2loopr}
\int\frac{d^d p}{(2\pi)^d}\frac{d^d q}{(2\pi)^d}
\frac{1}{p^2 q^4}
\end{eqnarray}
and
\begin{eqnarray}\label{eq:2loop}
\int\frac{d^d p}{(2\pi)^d}\frac{d^d q}{(2\pi)^d}
\frac{1}{p^2 q^2 (p+q)^2}
\end{eqnarray}
in $d$ dimension.
The Eq.~(\ref{eq:2loopr}) calculating with cut-off method
in the case of $d=4$ is divergent as $\Lambda^2\log(\Lambda^2/\mu^2)$.
When we use DREG,
the mass terms need to be considered to be associated with momentums in the denominator
of the formula to sidestep
the Infrared (IR) divergences, and the result
is proportional to $\Gamma(1-d/2)\Gamma(2-d/2)$. While
these kind of momentum integral can be safely neglected
when we consider quadratic divergences effects,
since these contributions are canceled on the basis of the one-loop
renormalization~\cite{CapdequiPeyranere:1990gk}.
The Eq.~(\ref{eq:2loop}) is what we should cares about,
which diverges as $\Lambda^2$
(when the higher energy cut-off($\Lambda$) is much larger than
the low energy cut-off scale($\mu$))when we use cut-off method
in the case of $d=4$,
and is proportional to $\Gamma(3-d)\Gamma(2-d/2)$
when using DREG method and including mass terms in the denominator of the formula.
Thus one have poles at $d=3,4$ on the complex $d$ plane.
The pole at $d=3$ corresponds to quadratic divergences
with the pole at $d=4$ corresponding to logarithmic divergences.

After we get the relation between quadratic divergences and
poles on the complex $d$ dimensional plane,
we need to relate quadratic divergences with logarithmic divergences
to reveal the physical meaning of the complex $d$ dimensional plane.
The DREG may keeps the physics of logarithmic
physics as $d\rightarrow 4$ which gives our traditional four dimensional
low energy physics.
With quadratic divergences effects($\Lambda^{2}$ relevant physics) being
expressed on the complex $4-2/L$ dimensional plane, parameters applied on which
might be expressed by the parameters of the $d=4$ case.

With loop order increasing by one, the quadratic divergences
contribution of quantum correction to mass term has
one more factor $1/16\pi^2$~\cite{Alsarhi:1991ji}
to multiply\footnote{The two-loop level case is
$\frac{1}{16\pi^2}\ln\frac{2^6}{3^3}$ times of that
of one-loop level case in the method adopt in~\cite{Hamada:2012bp}.},
thus the one-loop order's contribution dominates
the quadratic divergences.
Thus in the following paragraphs, we want to explore
the relation between the parameters
on the complex two dimensional plane and that of $d=4$ case.

\subsection{Relations between parameters
on the complex two dimensional plane and that of $d=4$ case}

At first, we consider how quadratic divergence
shows up with power counting method.
For any loop orders,
the momentum integral in $d=4-2/L$ dimension
Euclidean spacetimes which gives quadratic divergences
takes the form
 \begin{eqnarray}
 \int d^{4-2/L}k_E
 \frac{1}{(k_E^2+m^2)^{2-1/L}}\;,
 \end{eqnarray}
except other terms being multiplied to which to give constant terms at UV limit.
With the transformation
 \begin{eqnarray}\label{eq:km_trans}
 k_{E}\rightarrow k_{E}e^{k^2_{E}f(M)}\;,\quad
 m\rightarrow m e^{k^2_{E}f(M)}\;,
 \end{eqnarray}
where $f(M)$ take the responsibility to
compensate the dimension of interaction
couplings\footnote{The $f(M)$ can be specialized to
be $1/2M^{2}$ as will be shown in the one-loop level case,
and the inverse square of $M$ can be used
to compensate the mass dimension of interaction
couplings of the $d=4-2/L$ and $d=(4-2/L)+2$ cases.
In addition, the $f(M)$ needs to be proportional
to $M^{-2}$ to make the exponent in the transformation Eq.~(\ref{eq:km_trans}) dimensionless.}
and $M$ has mass dimension one,
we get the momentum integral
\begin{eqnarray}
\int dk_E
\frac{k_E^{5-2/L}}{(k_E^2+m^2)^{2-1/L}}\rightarrow\int dk_E
\frac{k_E^{3}}{(k_E^2+m^2)}\;,
\end{eqnarray}
for $k_E\gg m$,
which manifests quadratic divergences.

The reasonability of the above method to manifest quadratic divergences
can be traced back to the analytic
continuation property of the $\Gamma(x)$ function, with complex $x$.
The function reveals the pole after the definition region of
$\Gamma(x)$ has been analytic generalized
to the right of the pole on the complex $x$ plane to make the function
well defined~\cite{Leibbrandt:1975dj}.

For the $d=2$ case,
quantum corrections to the mass term at one-loop level
calculated with cut-off method
is
\begin{eqnarray}\label{eq:int2d}
\lambda\int\frac{d^{2}k}{(2\pi)^{2}}
\frac{1}{k^{2}-m^{2}}
=-i\frac{\lambda}{4\pi}\log\frac{\Lambda^{2}}{\mu^{2}}\;,
\end{eqnarray}
where the coupling $\lambda$ has mass dimensions $2$ based on
dimensional analysis for the missing of $\mu^{\varepsilon}$
to compensate the dimension of $\lambda$ compared
with Eq.~(\ref{eq:dreg}).
Firstly, let momentum and mass of the Eq.~(\ref{eq:int2d})
in Euclidean spacetime
take transformation as Eq.~(\ref{eq:km_trans}) with $f(M)=1/2M^2$,
then we get
 $d k^{2}_{E}\rightarrow e^{k^{2}_{E}/M^{2}}\big(dk^{2}_{E}
  +k^{2}_{E}dk^{2}_{E}/M^{2}\big)$.
Secondly, let the scalar quartic coupling transforms as
$\lambda\rightarrow\lambda M^{2}/(4\pi)$. After which,
the quantum correction to the mass term in Euclidean spacetime
 of $d=2$ case
\begin{eqnarray}\label{eq:int2d4dE}
\lambda\int\frac{d^{2}k_{E}}{(2\pi)^{2}}
\frac{1}{k_{E}^{2}+
m^{2}}
\rightarrow\frac{\lambda}{(4\pi)^2}
\big(\Lambda^{2}-\mu^{2}-m^{2}\log
\frac{\Lambda^{2}}{\mu^{2}}
\big)
+\frac{\lambda}{4\pi}\log\frac{\Lambda^{2}}{\mu^{2}}\;,
\end{eqnarray}
thus we arrive at a nontrivial result: the divergences
of $d=4$ case emerges from the $d=2$ case.
Changing back to Minkovski spacetimes,
no matter $\Lambda\gg\mu\gg m\gg M$ or $\Lambda\gg\mu\gg m$ with $M$
comparable to the low energy scale $\mu$,
we have
 \begin{eqnarray}
\lambda\int\frac{d^{2}k}{(2\pi)^{2}}
\frac{1}{k^{2}-m^{2}}
 \rightarrow -i\frac{\lambda}{(4\pi)^2}\big(\Lambda^{2}-\mu^{2}
\big)\;.
\end{eqnarray}
Where the quadratic divergences part of the $d=4$ case emerges from the
$d=2$ case, which supports to correspond the quadratic
divergences to the pole on the complex two dimensional plane.

Physical implications and discussions:
Considering the cut-off as the energy scale in the viewpoint of
effective field theory, we achieved that physics in the high-energy region of $d=4$
case can be described by that of the $d=2$ case. When the energy scale $\Lambda\gg\mu\gg m$,
the logarithmic divergences of $d=2$ case
gives rise to the quadratic divergences of the $d=4$ case.
Thus the quadratic divergences relevant physics up to one-loop level does can
live on the complex two dimensional plane as proposed by Veltman.
The logarithmic divergences can be considered as low energy physics compared
with quadratic divergences from
the viewpoint of effective field theory.
The $d=4$ case in DREG only preserves logarithmic divergences, thus the $d=4$ case
gives rise to low energy physics.
Since the logarithmic divergences is multiplicative renormalization
in the cut-off method and DREG, the $\lambda$ on the right hand side of
$\lambda\rightarrow\lambda M^{2}/(4\pi)$ is the $\lambda$ which is associated with
the renormalization of logarithmic divergences in the case of $d=4$ in DREG.
Thus the parameters on the complex two dimensional plane (physical parameters at high energy region)
can be connected
with that of $d=4$ case(low energy region), through which we can study high energy region physics
with parameters of the low energy region.

As one important application of the above arguments,
we derive the naturalness problem in $R_{\xi}$ gauge in the next section.
The RGE of the Higgs mass up to one-loop level on the complex two dimensional plane
will also be studied.
To derive the RGE of the Higgs mass,
the factor $M^2/4\pi$ associated with the $\lambda$ transformation needs to be replaced by a function of
the scale parameter $\mu$ in the $\rm{MS}$(or $\overline{\rm{MS}}$) scheme.

\section{Renormalization of the Higgs mass involving quadratic divergences}
\label{sec:rg2}

In order to extract
some useful physical consequences, it is suitable to explore RGEs of parameters of
the renormalizable theory in $\rm{MS}$ (or $\overline{\rm{MS}}$) scheme.
The scheme has the remarkable property that beta functions ($\beta$)
and anomalous dimension of mass ($\gamma_m$) derived in this scheme are all
gauge-independent~\cite{Caswell:1974cj,Gro:76}.
In other renormalization schemes, the renormalization coupling constant ($Z_{g,m}$)
for coupling or mass is gauge dependent in general for including finite terms in which.
For no explicit scale parameter ($\mu$) dependent in $Z_{g,m}$,
so RGEs of the couplings and $\gamma_{m}$ in $\rm{MS}$ (or $\overline{\rm{MS}}$)
scheme(which are functions of $Z_{g,m}$) carry no explicit $\mu$-dependent.
Thus $\rm{MS}$ (or $\overline{\rm{MS}}$) is always referred as the mass-independent
renormalization scheme. This property of $\rm{MS}$ (or $\overline{\rm{MS}}$) scheme
makes it easy to solve RGEs.
Based on the above argument, the $\rm{MS}$ (or $\overline{\rm{MS}}$) scheme will be chosen
as the renormalization scheme in this paper.

\subsection{Renormalization procedure and the VEV}

Considering the Higgs mass of the SM as one physical quantity,
the two-point connected Green function of the Higgs field
needs to be gauge invariant. When one study the perturbative correction
of the Higgs mass of the SM, one should take into account not only
standard loop corrections (the two point $\rm{1PI}$ self-energies),
but also corrections to the definition of VEV
via minima of the Higgs potential. The loop corrections to the definition of VEV,
entering through the so-called tadpole $\rm{\rm{1PI}}$ (one point $\rm{1PI}$ truncated Green function),
causes VEV shift.
The VEV shift induced by tadpole $\rm{1PI}$ values much in the mass renormalization,
which induced the $\rm{1PR}$ two-point self-energy of the Higgs field~\cite{Ma:1992bt},
with which we can get gauge invariant mass correction~\cite{Weinberg:1973ua,Fleischer:1980ub}.
This kind of renormalization method has also been
adopted in fermion mass renormalization procedure~\cite{Hempfling:1994ar}.
For the renormalization of other mass terms in the SM involving VEV,
the same procedure needs to be adopted in order that we get gauge invariant
masses.

Furthermore, the tadpole $\rm{1PI}$ is related with the Higgs potential~\cite{Coleman:1973jx}, with the tadpole $\rm{1PI}$ we can arrive at the
Higgs potential directly (see Eq.
$(3.20)$ and $(3.21)$ of the paper~\cite{Weinberg:1973ua}).
The paper~\cite{Lee:1974fj} derived the relation between
the tadpole $\rm{1PI}$ and the effective potential
in Landau gauge,
which makes the derivation of the Higgs potential technically
easier~\cite{Sher:1988mj}.

\subsection{The Lagarangian and the counter-term method}

Relations between renormalized masses and parameters used in this work are
\begin{eqnarray}\label{eq:notation}
 &&m_{H}=\sqrt{2\lambda}~v\;,\qquad m_{W}=\frac{g_{2}v}{2}\;,\qquad
   m_{Z}=\frac{g_{1}v}{2cos\theta_{W}}\;,\nonumber\\
 &&m_{t}=\frac{g_{t}v}{\sqrt{2}}\;, ~~\qquad {\rm cos}\theta_W=
 \frac{g_{2}}{\sqrt{g_{2}^{2}+g_{1}^{2}}}\;.
 \end{eqnarray}
with $\lambda$, $g_{t}$, $g_{2}$ and $g_{1}$ being scalar quartic coupling,
top quark Yukawa coupling, $SU(2)_L$ and $U(1)$ gauge couplings, respectively.

After spontaneous symmetry breaking, the bare Lagrangian of the
Higgs part of the SM in four dimensional spacetime is
\begin{eqnarray}
\mathcal{L}_{H}^{0}&=&\frac{1}{2}(\partial_{\mu}H^{0})^{2}
-\frac{1}{2}
(m^{0}_{H})^{2}(H^{0})^{2}-\lambda^{0}
v^{0}(H^{0})^{3}-\frac{1}{4}\lambda^{0} (H^{0})^{4} \nonumber\\
&& + \mathrm{const.}\;,
\end{eqnarray}
where superscripts $0$ on mass, couplings and the Higgs field stand for bare parameters.
Parameters which do not have superscripts represent renormalized parameters.
And $m^{0}_{H}=\sqrt{2\lambda^{0}}v^{0}$ has been set.
Let us first introduce four renormalization constants to relate bare parameters with renormalized ones,
\begin{eqnarray}\label{eq:rgec}
\lambda^{0}=Z^{-2}_{H}Z_{1}\lambda\;,\quad (m^{0}_{H})^{2}=Z^{-1}_{H}Z_{0}m_{H}^{2}\;,
\quad H^{0}=Z^{1/2}_{H}H\;.
\end{eqnarray}
Then, the relation between bare and renormalized VEV is given
by $v^{0}=Z^{-1/2}_{1}Z^{1/2}_{H}Z^{1/2}_{0}v$.
After renormalization constants being introduced, the bare Lagrangian of the SM, which can be used to extract physics of $d=4$ case, is recast as
\begin{eqnarray}
\mathcal{L}^{0}_{H}=\frac{1}{2}Z_{H}
(\partial_{\mu}H)^{2}-\frac{1}{2}Z_{0}m_{H}^{2}
H^{2}-Z^{1/2}_{1}Z^{1/2}_{0}\lambda
\mu^{\varepsilon}vH^{3}-\frac{1}{4}Z_{1}
\lambda\mu^{\varepsilon} H^{4}
 + \mathrm{const.}\;.
\end{eqnarray}
Where the arbitrary mass parameter $\mu$ is introduced
through $\lambda^{0}=Z_{\lambda}\lambda\mu^{\varepsilon}$
and $(g^{0})^{2}_{1,2,t}=Z_{g_{1},g_{2},g_{t}}g^{2}_{1,2,t}
\mu^{\varepsilon}$ with $\varepsilon=4-d$.
With $Z_{\lambda}$ and $Z_{g_{1},g_{2},g_{t}}$ containing poles at $d=4$,
beta functions of all couplings of the SM of $d=4$ case can be given easily
\footnote{It is customarily considered that the $d=4$ case in DREG corresponds
to four dimensional spacetime physics. While, according to our discussions in
section~\ref{sec:dim}, the $d=2$ case need to be included to describe high energy region
physics of four dimensional spacetime.}.

However, to derive RGE of the Higgs mass on the complex two dimensional plane,
the corresponding beta functions are needed.
These beta functions may be derived from the connections between
the parameters on the plane and the parameters of
the $d=4$ case, as will be explored bellow.
The bare Lagrangian of the SM, which can be used to extract the physics on
the complex two dimensional plane, takes the form of
 \begin{eqnarray}\label{eq:2dlag}
\mathcal{L'}^{0}_{H}=\frac{1}{2}Z'_{H}
(\partial_{\mu}H)^{2}-\frac{1}{2}Z'_{0}m_{H}^{2}
H^{2}-Z'^{1/2}_{1}Z'^{1/2}_{0}\lambda'\mu^{\epsilon}vH^{3}
-\frac{1}{4}Z'_{1}\lambda'\mu^{\epsilon} H^{4}
 + \mathrm{const.} \;.
\end{eqnarray}
Where the arbitrary mass parameter $\mu$ is introduced
through $\lambda^{'0}=Z'_{\lambda}\lambda'\mu^{\epsilon}$
and $(g^{\prime 0})^{2}_{1,2,t}=Z'_{g_{1},g_{2},g_{t}}g'^{2}_{1,2,t}\mu^{\epsilon}$ with $\epsilon=2-d$.
The superscript $\prime$ is to
identify the parameters on the complex two dimensional plane, these
parameters are related to parameters of $d=4$ case through $\lambda'=\lambda\mu^2$ and
$g^{\prime2}_{1,2,t}=g^{2}_{1,2,t}\mu^2$.
The bare Lagrangian of the Higgs part $\mathcal{L'}^{0}_{H}$ can be separated to be
renormalized
$\mathcal{L'}_{H}$ and the
counter-term $\mathcal{L'}^{ct}_{H}$ part, with $\mathcal{L'}_{H}$ precisely
equal to $\mathcal{L'}_{H}^{0}$
when bare parameters in $\mathcal{L'}_{H}^{0}$ are replaced by renormalized ones.
And the counter-term part is
\begin{eqnarray}\label{eq:lagarangian_ct}
\mathcal{L'}^{ct}_{H}=\frac{1}{2}(Z'_{H}-1)
(\partial_{\mu}H)^{2}-\frac{1}{2}(Z'_{0}-1)m_{H}^{2}
H^{2}-(Z'^{1/2}_{1}Z'^{1/2}_{0}-1)\lambda'\mu^{\epsilon} vH^{3}-\frac{1}{4}
(Z'_{1}-1)\lambda'\mu^{\epsilon} H^{4}
 + \mathrm{const.}\;,
\end{eqnarray}
with renormalization constants $Z'_{m}=Z'^{-1}_{H}Z'_{0}$ and $Z'_{\lambda}=Z'^{-2}_{H}Z'_{1}$.
We find that the beta functions on the complex
two dimensional plane can be expressed by
those of the $d=4$ case
\begin{eqnarray}\label{eq:b2d}
\mu\frac{d\lambda'}{d\mu}
=\mu^{2}\beta(\lambda(\mu))\;,\quad
\mu\frac{dg'_{1,2,3,t}}{d\mu}
=\mu\beta(g_{2}(\mu))\;.
\end{eqnarray}

\begin{figure}
  \centering
  \includegraphics[width=0.5\linewidth]{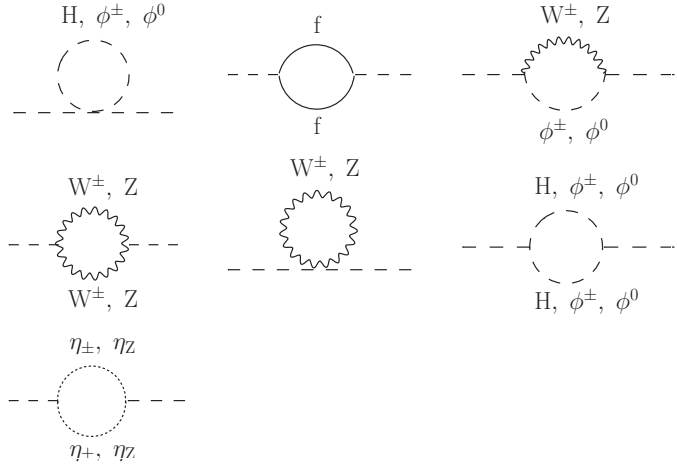}\\
  \caption{One-loop \rm{1PI} self-energy corrections to the Higgs mass. }\label{Fig:Higgs1PI}
\end{figure}

\begin{figure}
  \includegraphics[width=0.5\linewidth]{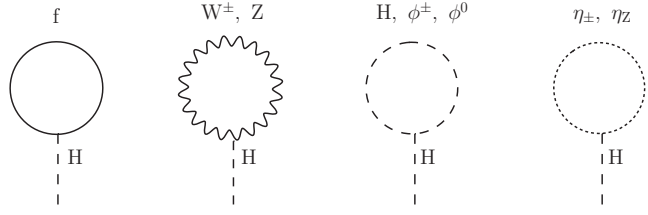}\\
  \caption{Tadpole Feynman diagrams with one external Higgs field in the SM. }\label{fig:tad1loop}
\end{figure}

In $\rm{MS}$ (or $\overline{\rm{MS}}$) scheme, divergent terms in self-energy
of the Higgs field can be subtracted through renormalization constants.
Two-point self-energy of the Higgs field comes from $\rm{1PI}$ self-energy and tadpole
contributions
\begin{eqnarray}\label{eq:QD_IPI}
\Sigma_{H}(p^{2})=\Sigma_{H}^{\rm{\rm{1PI}}}(p^{2})
+\Sigma_{H}^{T}(p^{2})\;.
\end{eqnarray}
Feynman diagrams which contribute to $\rm{1PI}$ self-energy are shown in Fig.~\ref{Fig:Higgs1PI}.
The tadpole diagrams contribution to the self-energy of the Higgs boson is
\begin{equation}
\Sigma_{H}^{T}=-i\frac{3m_{H}^{2}}{v}
\frac{i}{-m_{H}^{2}}T\;,
\end{equation}
where $i/(-m_{H}^{2})$ is the propagator of the Higgs boson carrying
zero momentum, and the Higgs three-point vertex is $-i3m_{H}^{2}/v$,
with amplitude $T$ being depicted in Fig.~\ref{fig:tad1loop}.
Up to one-loop level, the counter-term method requires
$\Sigma_{H}(p^{2})+i(Z'_{H}-1)p^{2}-i(Z'_{0}-1)m^{2}_{H}=0$. Combining with
the relation $(m^{0}_{H})^{2}=Z'_{m}m_{H}^{2}$, we derive the renormalization
constant of the Higgs mass ($Z'_{m}$) on the complex two dimensional plane
in next subsection.

\subsection{Renormalization constant and anomalous dimension of the Higgs mass}
\label{sec:rg}

Proceeding calculations on the complex two dimensional plane with DREG,
the renormalization
constant for the Higgs field is obtained to be $Z'_{H}=1$ for the absence of poles at $d=2$.
The renormalization constant
$Z'_{0}$ includes contributions of poles at $d=2$ and equals to
\begin{eqnarray}
Z'_{0}=1+\frac{2}{(4\pi)m^{2}_{H}}\frac{1}{1-d/2}
\big[6\lambda'-\frac{3}{2}Tr[I]g'^{2}_{t}
+(g^{\mu}_{\mu}-1)(\frac{3g'^{2}_{2}}{4}
+\frac{g'^{2}_{1}}{4})\big]\;.
\end{eqnarray}
Therefore, the renormalization constant of the Higgs mass is
\begin{eqnarray}
\label{eq:Zm'}
Z'_{m}=1+\frac{2}{(4\pi)m_{H}^{2}}\frac{1}{1-d/2}\big[6\lambda'
-\frac{3}{2}Tr[I]g'^{2}_{t}
+(g^{\mu}_{\mu}-1)(\frac{3g'^{2}_{2}}{4}
+\frac{g'^{2}_{1}}{4})\big]\;.
\end{eqnarray}

From Eq.~(\ref{eq:Zm'}), it is easy to find that,
after one take Tr[I]=$g^{\mu}_{\mu}$=4 and the replacement
$1/(1-d/2)\rightarrow\Lambda^{2}/(4\pi)$,  the
naturalness problem can be expressed by
\begin{equation}\label{eq:DRED}
(m^{0}_{H})^{2}=m_{H}^{2}+\frac{2\Lambda^{2}}
{(4\pi)^{2}v^{2}}\big[3m_{H}^{2}-12m_{t}^{2}
+6m_{W}^{2}+3m_{Z}^{2}\big]
\end{equation}
in $R_\xi$ gauge.
Thus the naturalness problem is gauge independent.
Here one may argue that the naturalness problem is the appearance
of physics on the complex two dimensional plane.
And from Eq.~(\ref{eq:DRED}), one may easily find that the naturalness problem disappears
when the Veltman condition~\cite{Veltman:1980mj,Decker:1979cw} is satisfied, which is the key of SUSY
\footnote{For no sign of SUSY being observed at LHC so far, if SUSY exists,
it might be realized at high enough scale at which the Veltman condition is satisfied.
And the matching between the supersymmetric and non-supersymmetric theories should be done at that scale~\cite{Masina:2013wja}.}.

From the above arguments, any attempt to analysis
naturalness problem with RGEs of the SM which applied to $d=4$ case
may not be appropriate, RGEs that can be
applied to the complex two dimensional
plane are needed.
With renormalization constant of the Higgs mass term on the plane, the RGE of the Higgs mass can
be derived. The behavior of the Higgs mass on the plane with respect
to the energy scale $\mu$ can be treated as the effect caused
by the naturalness problem, i.e.,
the $\rm{MS}$ (or $\overline{\rm{MS}}$) substraction procedure keeps
the structure of the pole terms which is the source of the naturalness problem.

For the bare Higgs mass is independent of $\mu$,
RGE of the Higgs mass on the complex two dimensional plane can be derived
\begin{eqnarray}\label{eq:gamH}
\mu\frac{dm^2_{H}}{d\mu}&=&-m^2_{H}\lim_{\epsilon\rightarrow 0}
\gamma^{\prime}_{m^2_{H}}(m_{H}(\mu),\epsilon)\;,
\end{eqnarray}
where
\begin{eqnarray}
\gamma'_{m^2_{H}}(m_{H}(\mu),\epsilon)
=\frac{\mu}{Z'_{m}}\left(\frac{\partial Z'_{m}}{\partial \lambda'}
\beta(\lambda'(\mu),\epsilon)+\frac{\partial Z'_{m}}{\partial g'_{t}}
\beta(g'_{t}(\mu),\epsilon)+\frac{\partial Z'_{m}}{\partial g'_{1}}
\beta(g'_{1}(\mu),\epsilon)+\frac{\partial Z'_{m}}{\partial g'_{2}}
\beta(g'_{2}(\mu),\epsilon)\right)
\end{eqnarray}
represents anomalous dimension of the Higgs mass term.
Beta functions on the plane can be
expressed through beta functions of $d=4$ case as shown by Eq.~(\ref{eq:b2d}).
Since the anomalous dimension of the Higgs mass
is the function of $Z'_{m}$, it must be gauge independent. While in other
renormalization schemes the renormalization constant of the Higgs mass term is gauge dependent
in general. This is caused by appearance of the finite terms
in addition to the terms given in $Z'_{m}$ on the right hand side of Eq.~(\ref{eq:Zm'})
in other schemes, which is the same with the situation discussed at the begining of this section.
After one careful calculation, the anomalous dimension of the Higgs mass term
on the plane is derived
\begin{eqnarray}\label{eq:gam2d}
\gamma^{\prime}_{m^2_{H}}(m_{H}(\mu))=-\frac{1}{(4\pi)m^{2}_{H}}
(24\lambda'-12g'^{2}_{t}+g'^{2}_{1}+3g'^{2}_{2})\;.
\end{eqnarray}
Obviously, this equation is gauge independent, and we need to point out
that the occurrence of $m^{2}_{H}$ in the above equation is caused by the
form of the pole at $d=2$. While this equation is still dimensionless since
couplings in the parentheses on the right hand side of this equation take
mass dimension $2$. In the derivation of above equation, $Tr[I]=2$
and $g^{\mu}_{\mu}=2$ have been adopt based on DREG
argument~\cite{CapdequiPeyranere:1990gk}.
With the relations between parameters of $d=2$ and $d=4$ cases,
being given under Eq.~(\ref{eq:2dlag}), Eq.~(\ref{eq:gam2d})
can be represented by the parameters of $d=4$ case,
\begin{eqnarray}\label{eq:gam2d4}
\gamma^{\prime}_{m^2_{H}}=-\frac{\mu^{2}}{(4\pi)m^{2}_{H}}
(24\lambda-12g^{2}_{t}+g^{2}_{1}+3g^{2}_{2})\;.
\end{eqnarray}

While based on Veltman's argument on the freedom of vector bosons,
the $g_\mu^\mu$ and $Tr[I]$ all need to equals to $4$~\cite{Veltman:1980mj}.
Thus one can connect the RGE of the Higgs mass on the complex
two dimensional plane with the Veltman condition
\begin{eqnarray}\label{eq:velt}
\mu\frac{dm^2_{H}}{d\mu}=\frac{2\mu^{2}}{4\pi}VC(\mu)\;,
\end{eqnarray}
with
\begin{eqnarray}
VC(\mu)=12\lambda-12g^{2}_{t}+\frac{3g^{2}_{1}+9g^{2}_{2}}{2}\;,
\end{eqnarray}
and $VC(\mu)$ equaling to zero is the Veltman condition.
Now, we can use Eq.~(\ref{eq:velt}) together with beta functions of
couplings of the SM of $d=4$ case to study the physics
on the complex two dimensional plane.

\subsection{Scale-dependent property of the Higgs mass
on the complex two dimensional plane}
\label{sec:mH2d}

To explore the $\mu$-dependent property of the Higgs mass on the complex two dimension plane,
all RGEs of couplings of the SM are needed.
These beta functions up to two-loop order is listed
in Appendix~\ref{sec:beta}.
Boundary conditions of $g_{1,2,t}$ can be obtained as in~\cite{Holthausen:2011aa},
and the boundary conditions of $m_{H}$ is set to be $m_{H}($126 GeV)$=126$ GeV.
Considering the theoretical error in derivation of the top quark pole mass from the running one,
we choose the top quark pole mass $M_t=173.3\pm2.8$ GeV~\cite{Alekhin:2012py}, which strongly
affects the behaviors of RGEs. And other input parameters
are chosen to be the central values~\cite{J. Beringer}.
\begin{figure}[!htb]
  \centering
  \includegraphics[width=3.5in]{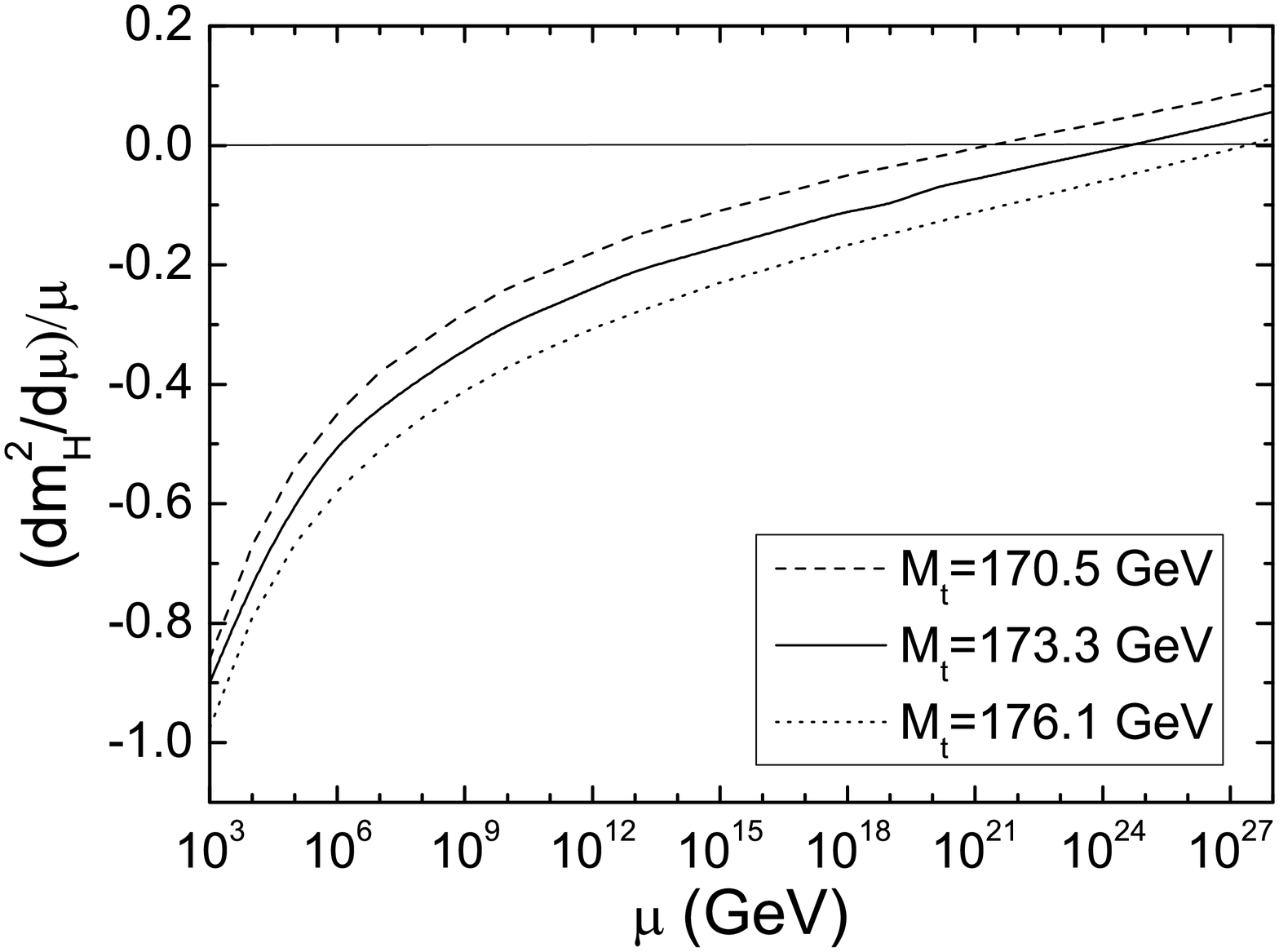}
  \centering
  \includegraphics[width=3.5in]{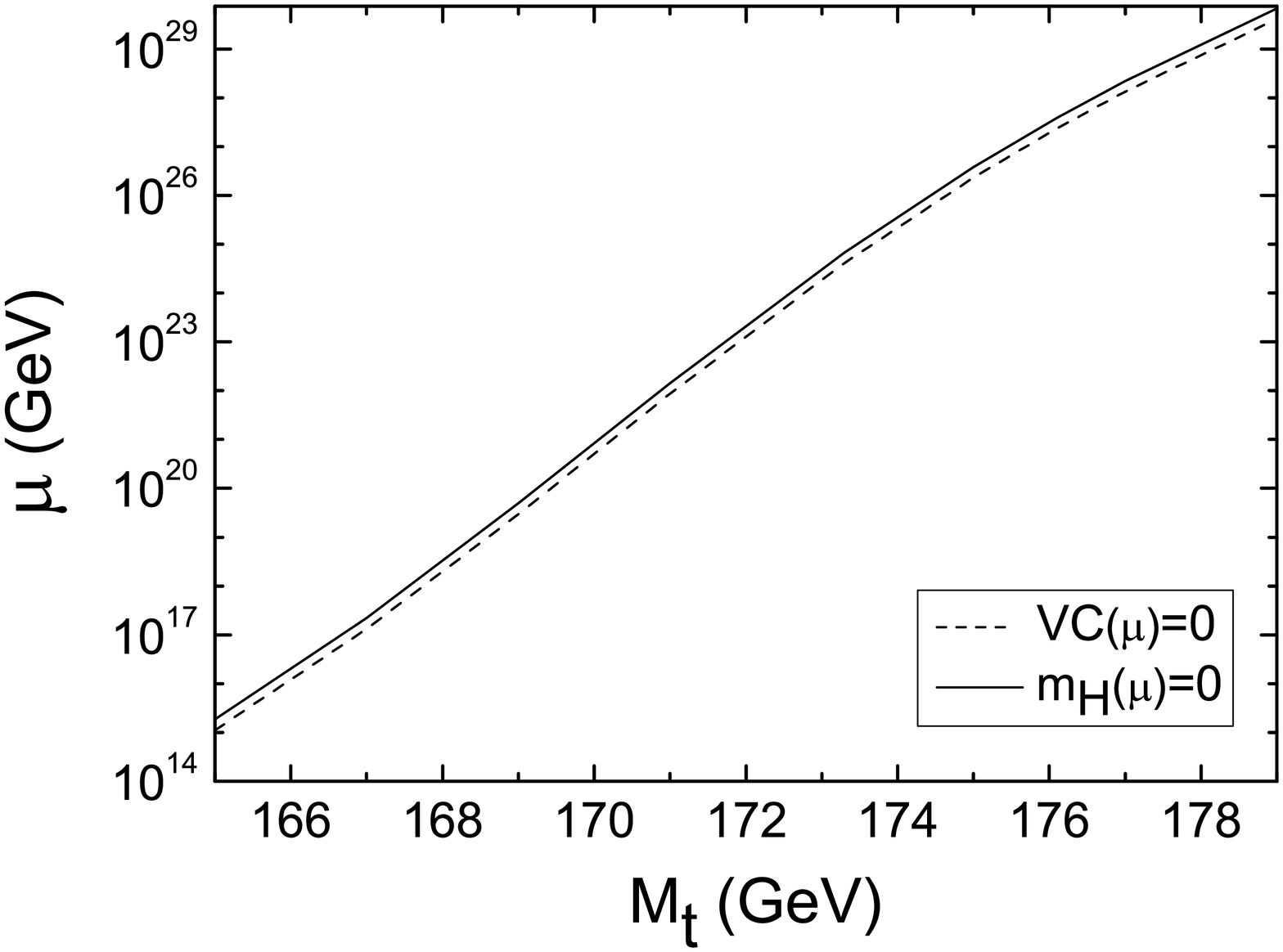}\\
  \caption{Left:Behaviour of the RGE of $m^2_{H}$ in unit of energy scale ($\mu$) on the complex two dimensional plane. Right:The UV scale at which $m_{H}=0$ with respect to $M_t$.}\label{fig:mH2d}
\end{figure}

From the behavior of RGE of the $m^{2}_{H}$ as depicted
on the left panel of Fig.~\ref{fig:mH2d}, one expect $m^{2}_{H}$ first decreasing and
latter increasing for the change of the sign of
$d m^2_{H}/d\mu$ with the energy scale increasing\footnote{That is caused by contributions which are proportional of the
couplings ($\lambda,~g_{1,2,t}$) in $\gamma'_{m_{H}}$ changes with the energy scale increasing,
i.e., contributions of $\lambda,~g_{1,2}$ become bigger than that of $g_{t}$, as shown
in~\cite{Masina:2013wja,Jegerlehner:2013cta}.}.
The $|m^2_H|$ running to be very large between the Electroweak scale and the Planck scale
\footnote{Different form the paper \cite{Hamada:2012bp}, we
work in the broken phase, where the gauge invariant property of $m_{H}$ can be
protected as discussed in previous
subsections. After we renormalized away the quadratic divergences induced by tadpole diagrams,
we have negative corrections to the $m^2_{H}$, that's why we use $|m^2_H|$ but $m^2_{H}$ here.
And in symmetric phase, one impose
$\langle H\rangle=0$, the corrections to the $m^2_{H}$ will be positive after one
renormalized away quadratic divergences. Take one-loop corrections as an example, only
$\Sigma_{H}^{\rm{1PI}}(p^{2})$ contribute to $m^2_{H}$
in symmetric phase and the $\Sigma_{H}^{T}(p^{2})$ needed to be considered in Eq.~($\ref{eq:QD_IPI}$),
thus Eq.~($\ref{eq:velt}$) changes to be $\mu dm^2_{H}/d\mu=-(\mu^{2}/4\pi)VC(\mu)$.
And the structure of quadratic divergences in symmetric phase and broken phase are the same, one
can expect the UV scales where $m_{H}=0$ are the same in the two cases.},
because in that energy region $(d m^2_{H}/d\mu)/\mu$ (connect with $VC(\mu)$
through Eq.~($\ref{eq:velt}$)) is always negative.
In fact, this property of the $m^2_H$ on the plane is caused by the naturalness problem, i.e.,
the quadratic divergences which cause the naturalness problem
are also multiplicatively renormalized as explored by us, thus the effect caused by naturalness problem
can be studied through the RGE of the $m^2_{H}$ on the plane.

The $(d m^2_{H}/d\mu)/\mu$(which is proportional to $VC(\mu)$) approaches zero at UV scale,
as is plotted on the left panel of Fig.~\ref{fig:mH2d}.
After $VC(\mu)$($(d m^2_{H}/d\mu)/\mu$) changes sign, the $m^2_{H}$ approaches
one nontrival zero quickly, the energy
scale which satisfies $m_{H}=0$ is slightly higher than the scale at which $VC=0$,
as is depicted on the right panel of Fig.~\ref{fig:mH2d}.
However, after $m^2_{H}=0$ being achieved, with the increase of energy scale, $m^2_{H}$ can still get large value
due to the sign-flip of $(d m^2_{H}/d\mu)/\mu$, which is also noted in~\cite{Jegerlehner:2013cta}.
This behavior can be explained by the fact that the roles played by the higher energy scale
$\Lambda$ where the effective theory can be applied, and the lower energy scale $\mu$ shown in
Eq.~(\ref{eq:cutoff}) are interchanged.
Above argument is also suitable for our renormalization procedure.

The dependence of energy scale at which $m^2_{H}(VC)=0$
on the top quark pole mass, is plotted on the right panel of Fig.~\ref{fig:mH2d}.
Assuming the SM can be valid to arbitrarily large energy scale,
one can always expect the value of $m_{H}$ to arrive at zero.
And $m^2_{H}=0$ will be achieved at about the Planck scale
for $M_t=168$ GeV, which is about $2\sigma$ smaller
than the central value.
The scale increases with the value of $M_{t}$ increasing.
For $M_{t}$=173.3 GeV, the scale is around $10^{24}$ GeV. And for the scale exceeding
the Planck scale, it was extensively believed that the gravitational contributions
need to be considered~\cite{Hamada:2012bp}.

For the multiplicative renormalization property
of the Higgs mass-square on the complex two dimensional plane,
one can have $m^0_{H}=0$ when $m_{H}=0$. And the vanishing of the bare Higgs mass
is the main result of the paper~\cite{Hamada:2012bp},
where it was argued to hint the restoration of SUSY.
While the paper~\cite{Masina:2013wja} argued that
the SUSY can match with the SM at the scale $VC=0$. Based on our
analysis, we find that one can not expect $m^0_{H}=0$ and $VC=0$ to be
satisfied at the same scale.

Based on Bardeen's argument on the naturalness problem, quadratic divergences could be
safely removed by imposing boundary condition at the UV
(Planck) scale $M_{pl}$~\cite{Bardeen:1995kv,Iso:2012jn,Iso:2013aqa}, and the behavior
of $m^2_{H}$ plotted in Fig.~\ref{fig:mH2d} suggests the natural choice could be $m_{H}(M_{pl})=0$
\footnote{Based on argument of the paper~\cite{Iso:2013aqa}, the natural boundary condition of
the mass term at the $M_{pl}$ is chosen to be $m^{2}(M_{pl})=0$, with $m^2$ is the
mass term in the tree level Higgs potential with the `wrong' sign, which is
equivalent to our result for $m^2_{H}=-2m^2$ after considering the SSB.
Here, one need to note that the condition $m_{H}(M_{pl})=0$ as was explored by us
can be independent of the matter content~\cite{Iso:2013aqa}, which is different from the
approaching zero of the Veltman condition which depends on
the matter content.}.
Imposing this condition on the RGE of $m_{H}$ for $d=4$ case within DREG,
which is gauge invariant
and multiplicatively renormalized result of logarithmic divergences,
we always have one zero mass,
with implications: Attributing the naturalness problem to the physics on the
complex two dimensional plane, RGEs of $d=4$ case within DREG method can be
described by logarithmic terms safely.
And paper~\cite{Iso:2012jn,Iso:2013aqa} supposes the boundary condition
imposed on the SM need to be justified in the UV complete theory.
The solution of
the naturalness problem calls for knowledge of UV complete theory~\cite{Masina:2013wja}.
On the complex two dimensional plane, we can study quadratic divergences effect
in a gauge invariant way, and the quadratic divergences correspond to
UV physics in the viewpoint of effective field theory.
With the naturalness problem lives on the complex two dimensional plane,
to solve the problem, one need to investigate the property of the complex
two dimensional plane more.

If we study the RGE of the $m^2_{H}$ described by
Eq.~($\ref{eq:gam2d4}$) not Eq.~($\ref{eq:velt}$) on the complex two dimensional plane,
then the UV scale where $m_{H}$
vanishes will be improved slightly, with other related discussions will not be changed.

\section{The viewpoint of Wilsonian renormalization group on quadratic divergences}
\label{sec:wilRG}

In this section, We study the role played by quadratic divergences
in the Wilsonian renormalization group method.

From the viewpoint of Wilsonian renormalization group,
when the energy scale $\Lambda\gg m$, the correction to the mass term
of $\lambda\phi^{4}$ theory,
came from quantum contributions of the energy region $b\Lambda\leq|k_{E}|\leq \Lambda$~\cite{Peskin:1995ev}
\begin{eqnarray}\label{eq:wilson}
\frac{1}{2}\lambda\int\frac{d^{d}k_{E}}{(2\pi)^{d}}
\frac{1}{k_{E}^{2}}=
\frac{1}{(4\pi)^{d/2}\Gamma(d/2)}\frac{(1-b^{d-2})}{d-2}
\Lambda^{d-2}\;,
\end{eqnarray}
where the subscript $E$ denotes the parameters living in Euclidean spacetime.
In the $d$ dimensional $\lambda\phi^{4}$ theory, the relation between masses at energy
scale $b^{n}\Lambda$ (m) and
$\Lambda$ ($m'$), with $b<1$ but very close to $1$, is given by
\begin{eqnarray}\label{eq:masswilson}
m'^{2}=m^{2}b^{-2n}+\frac{\lambda\pi^{d/2}}
{\Gamma(d/2)(2\pi)^{d}}
\frac{1-b^{d-2}}{d-2}\Lambda^{d-2}
\sum^{n}_{n=1}b^{(d-2)n-d}\;,
\end{eqnarray}
where quantum corrections to $\phi$ are dropped for small contributions.
When dimension $d$ is compacted to $4$ through continuation, the mass relation is
 \begin{eqnarray}\label{eq:wilmass}
 m'^{2}=m^{2}b^{-2n}+\frac{n\lambda(b^{-2}-1)
 \Lambda^{2}}{32\pi^{2}}\;,
 \end{eqnarray}
with the $n$ in the above two equations being constraint by the condition $\Lambda\sim m'$.
After iterating $n$ times the Lagrangian from scale $b^{n}\Lambda$ to $\Lambda$,
as shown in~\cite{Peskin:1995ev},
the difference between $m'^{2}$ and $m^{2}$ comes to be enormous, which calls for
delicate choosing of $m'^{2}$ when one wants to know the $m^{2}$ at one lower momentum
scale $b^{k}\Lambda$,
with $b^{n}\Lambda\leq b^{k}\Lambda<\Lambda$. We would like to point out that
the quadratic divergences
can be subtracted through one new appropriate choice of coordinates of the theory space,
i.e., with which being absorbed into
$m'^{2}_{new}$.
Therefore, the difference between $m^{2}$ and $m'^{2}_{new}$ is fully from rescaling of distance
and the field $\phi$~\cite{Peskin:1995ev}, i.e., $m^{2}=m'^{2}_{new}b^{2k}$.
The real issue of the naturalness problem is not the quadratic divergences
but the rescaling effects.
In addition, the operation of absorbtion of the $\Lambda^{2}$ terms in the above argument
is indeed the attribution of quadratic divergences to the physics on the complex two dimensional plane
following our discussions in section~\ref{sec:dim}.
Furthermore, quadratic divergences up to all loop orders could be absorbed to give
the critical surface in the sense of Wilsonian group method~\cite{Aoki:2012xs},
this kind of surface plays the role similar with the complex $4-2/L$ dimensional plane.
After the quadratic divergences being subtracted, one new coordinate of space of parameters
are given, and the RG flows around the critical surface is determined by logarithmic divergences.
With DREG, all the four dimensional physics, i.e., RGEs of physical parameters
may be achieved at $d\rightarrow 4$, which corresponds to logarithmic divergences.
With quadratic divergences living on the complex $4-2/L$ plane,
which would not change our $d=4$ physics, one can separate quadratic divergences
and multiplicative logarithmic divergences
safely~\cite{Bollini:1972ui,'tHooft:1972fi,Fujikawa:2011zf,Aoki:2012xs},
though they are all multiplicatively renormalized as explored in last two sections.
When the dimensionality $d$ is compacted to $2$, mass correction
Eq.~(\ref{eq:masswilson}) gives rise to $-\lambda\log b/(4\pi)$. Thus
$m'^{2}=m^{2}+\lambda\log b/(4\pi)$. The variance of $m$ is not so
rapid now, and the RGE of mass
can be derived
\begin{eqnarray}
\frac{d(m'^{2}-m^2)}{d\log b}=\frac{\lambda}{4\pi}\;.
\end{eqnarray}
Generalizing to the SM case, the RGE of the Higgs mass can be computed
directly,
and the same result with that derived in section~\ref{sec:rg} can be obtained
considering the correspondence between $b$ and the $\mu$.

For the case of scalar quartic coupling $\lambda$ in the $\lambda\phi^{4}$ theory,
the relation between quantities at the scale $\Lambda$ and $b^{n}\Lambda$ is given by
\begin{eqnarray}
\lambda'=\left(\lambda b^{(d-4)n}-\frac{3\lambda^{2}}{16\pi^{2}}
\log\frac{1}{b}\sum^{n}_{n=1}b^{(d-4)n}\right)\;.
\end{eqnarray}
Apparently, when dimension $d$ is continued to $4$, this formula will be
greatly simplified,
\begin{eqnarray}\label{eq:lamcorwil}
\lambda'=\left(\lambda-\frac{3\lambda^{2}}{16\pi^{2}}
\log\frac{1}{b}\right)\;.
\end{eqnarray}
Where no $n$ shows up as in Eq.~(\ref{eq:wilmass}). Indeed, this is caused
by the fact that the mass term $m^{2}\phi^{2}$ and the scalar
quartic interaction $\lambda\phi^{4}/4$ are relevant
and marginal terms respectively in four dimension.
The $\lambda$ does not have the initial value choosing problem
as the mass term when we get its
value at one energy scale lower than $\Lambda$.
When one study the quantum corrections to $\lambda$ in the $\lambda\phi^{4}$ theory
in four dimensional spacetime, only the correspondence
between the pole at $d=4$ and logarithmic divergences(the second formula of Eq.~(\ref{eq:correspondence}))
can be found to any loop orders.
Thus, the quantum corrections of $\lambda$ is proportional to
$\log(\Lambda^{2}/\mu^{2})$, which corresponds to the case
of Eq.~(\ref{eq:lamcorwil}) when the scale runs from $\mu$ to $b^{n}\Lambda$.
In the ordinary perturbation calculation as in section~\ref{sec:rg2},
the quantum corrections to $\lambda$ is
wholly the same with Eq.~(\ref{eq:lamcorwil}) when we take $\mu=b^{n}\Lambda$,
which is due to the rescaling distance effect
disappearing in four dimensional spacetime, i.e., $b^{(d-4)n}\rightarrow b^{0}=1$
as $d\rightarrow 4$. We should mention that couplings($\lambda,~g_{1,2,3,t}$) of the
SM also share the same property as been discussed above.

\section{VEV and the composed Higgs mass}
\label{sec:cmH}

The rescaling property of the Higgs mass is different from that
of the Higgs quartic coupling since it still takes the rescaling
factor $b$ as $d\rightarrow 4$.
The rescaling distance effect in Eq.~(\ref{eq:wilmass}) calls for
delicate choosing of the mass of the theory at the scale $\Lambda$,
the core of the naturalness problem is the rescaling effect in the
Wilsonian sense.
In this section, we consider the opposite scenario in which the Higgs mass term
has the same rescaling property as that of the Higgs field in the SM,
and show that
the naturalness problem will not come to us in this case. Then we
study RG behaviors of physical parameters up to UV scale.

\subsection{RGE of VEV}

The SM has achieved great success, almost fits all experimental
results in last decades.
In the SM, the Higgs mechanism provides masses to fermions and gauge bosons, and one proper aim of
LHC is to check this mechanism. One SM-like Higgs signal with mass about
$126$ GeV has already been found at LHC~\cite{Aad:2012tfa}.
In this paper, we suppose that
the signal is just the Higgs of the SM. How to understand the mass term properly is
the key to understand the Higgs boson and the naturalness
problem. In the following paragraphs, we will show that the naturalness problem
does not shows up if the Higgs mass is considered not to be one gauge invariant but one composed quantity.

The key of the Higgs mechanism is the existence of the mass term which has the ``wrong" sign,
and can be given by\footnote{The $\mu^2$ is used to make the statement in this part clear, which should be
different from the scale parameter introduced in the $\rm{MS}(\overline{\rm{MS}})$ scheme. }:
$\mu^{2}=\lambda\phi^{2}|_{\phi=\phi^{0}} $,
with the $\phi^{0}$ being the value of $\phi$
where the minimum of the potential $V(\phi)$ occurs.
Thus the mass term(after SSB) $m^{2}$ can be given by $2\lambda\phi^{2}|_{\phi=\phi^{0}}$,
which is just two times of the product of $\lambda$
and $\phi^{2}$ when $\phi=\phi^{0}$,
we refer the $m^{2}$ as the composed mass.
Hereafter, we analyze property of the $m^{2}$ in Wilsonian sense
and do not view the $m^{2}$ as the value
with mass dimension two directly.
Firstly, imposing the parameters in the Lagrangian
at the energy scale $\Lambda$ takes the superscript $\prime$ and the
the Lagrangian
at the energy scale $b^{n}\Lambda$ does not.
The mass term of the Lagrangian at the energy scale $\Lambda$ is
$2\lambda'(\phi'^{0})^{2}\phi'^{2}$ with the $\phi'^{0}$ is the value
of $\phi'$ at the minima of $V(\phi')$.
And, the mass term at the energy scale $b^{n}\Lambda$ is
$2\lambda(\phi^{0})^{2}\phi^{2}$ with $\phi^{0}$ having the same
scaling property as $\phi$.
Thus the relationship between $\phi^{0}$ and $\phi'^{0}$ is the same as that
of $\phi$ and $\phi'$, and the relation
$\phi'^{0}=[b^{2-d}(1+\delta Z)]^{1/2}\phi^{0}$ holds when we iterating
the Lagrangian from energy scale
$b\Lambda$ to $\Lambda$.
Since we do not consider the $2\lambda\phi^{2}|_{\phi=\phi^{0}}$ as
the mass of field in the Lagrangian directly, we will not
encounter the enormous fine-tuning. The composed mass term
at the energy scale $\Lambda$ can be given by
$2(\lambda+\delta\lambda)(\phi^{0})^{2}\phi^{2}$ when $d\rightarrow 4$.
Take RG transformation $n$ times, i.e., successive iterate the transformation
procedure on the Lagrangian like from b$\Lambda$ to $\Lambda$ for $n$ times,
the relation between the composed mass terms at the energy scale $\Lambda$
and $b^{n}\Lambda$ can be given by
\begin{eqnarray}\label{eq:mwils}
2\lambda'(\phi'^{0})^2\phi'^{2}&=&2(\lambda+\delta\lambda)
b^{(d-4)n}(\phi'^{0})^{2}\phi'^{2}\nonumber\\
&\sim&2(\lambda+\delta\lambda)(\phi^{0})^{2}\phi^{2}, ~\rm{when}~ d\rightarrow 4\;,
\end{eqnarray}
with the quantities following the symbol ``$\sim$" living in energy scale
$b^{n}\Lambda$, and where the $n$ is absent for the same reason as argued in the last section,
and the effects caused by $\delta Z$ have been dropped for small contributions.

In the SM, the $\phi^{0}$ is the VEV, i.e., $v$, which shares the same scaling property as
the Higgs field. Translate the Eq.~({\ref{eq:mwils}) to the usual renormalization method,
wherein RGEs are calculated with
DREG based on $\rm{MS}$ (or $\overline{\rm{MS}}$) scheme, we have
\begin{eqnarray}
\lambda v^{2}\mu^{\varepsilon}Z_{1}Z^{-1}_{H}=\lambda_{0}v^{2}_{0}\;,
\end{eqnarray}
where the $v^{0}$ and $\lambda_{0}$ are the correspondences of $\phi'^{0}$
and $\lambda'$ respectively. Now the Higgs mass in the SM
should be considered as one composed mass.
Since $\lambda_{0}v^{2}_{0}$ does not depend on energy scale $\mu$
introduced in DREG and $\rm{MS}$ (or $\overline{\rm{MS}}$) scheme, we can derive the
RGE of $\lambda v^{2}$ in our four dimensional spacetime,
\begin{eqnarray}\label{eq:lavev1}
\mu\frac{d}{d\mu}(\lambda v^{2})=-2\lambda v^2 \gamma_{H}+v^2\beta_{\lambda}\;,
\end{eqnarray}
with one-loop anomalous dimension of the Higgs field is
\begin{eqnarray}\label{eq:gamHl}
\gamma^{(1)}_{H}=\frac{1}{64\pi^2}
(12g^{2}_{t}-9g^{2}_{2}-3g^{2}_{1})\;,
\end{eqnarray}
and the two-loop level contributions to $\gamma_{H}$~\cite{Einhorn:1992um} is
\begin{eqnarray}\label{eq:gamH2}
\gamma^{(2)}_{H}&=&\frac{1}{(16\pi^2)^2}
\big(6\lambda^{2}-\frac{27}{4}
g^{4}_{t}+20g^{2}_{3}g^{2}_{t}
+\frac{45}{8}g^{2}_{2}g^{2}_{t}
+\frac{85}{24}g^{2}_{1}g^{2}_{t}
-\frac{271}{32}g^{4}_{2}+\frac{9}{16}
g^{2}_{1}g^{2}_{2}+\frac{431}{96}g^{4}_{1}\big)
\end{eqnarray}
in Landau gauge, this gauge is appropriate in the sense of~\cite{Nielsen:1975fs}
based on effective potential argument.
And $\beta_{\lambda}$ is given in~\cite{Holthausen:2011aa}.
We can now isolate the RGE of VEV from Eq.~(\ref{eq:lavev1}),
\begin{eqnarray}\label{eq:advev}
\mu\frac{d}{d\mu}v^{2}=-2v^{2}\gamma_{H}\;,
\end{eqnarray}
with $\gamma_{H}$ are given by Eq.~(\ref{eq:gamHl},\ref{eq:gamH2}) at one-loop
and two-loop order, and coincide with the anomalous dimension of VEV
derived in~\cite{Blumhofer:1992ip,Arason:1991ic} at one- and two- loop order.
We want to note that this approach does sidestep the
naturalness problem, while the gauge invariant property of the RGE of the
composed Higgs mass (still denoted as $m^{2}_{H}=2\lambda v^{2}$ but with
different meaning with the $m^{2}_{H}$ considered in section~\ref{sec:rg2}) disappears,
since
 \begin{eqnarray}
 \mu\frac{dm^{2}_{H}}{d\mu}=2\mu\frac{d}{d\mu}(\lambda v^{2})
 \end{eqnarray}
and the RGE of the VEV, which is shown in Eq.~(\ref{eq:advev}), depends on the gauge parameter.

\subsection{Scaling property of the composed Higgs mass}

With the RGE of $v^{2}$(Eq.~(\ref{eq:advev})), we can study the behavior
of the VEV with respect to the energy scale $\mu$.
\begin{figure}[!htp]
  \centering
  \includegraphics[width=0.6\textwidth]{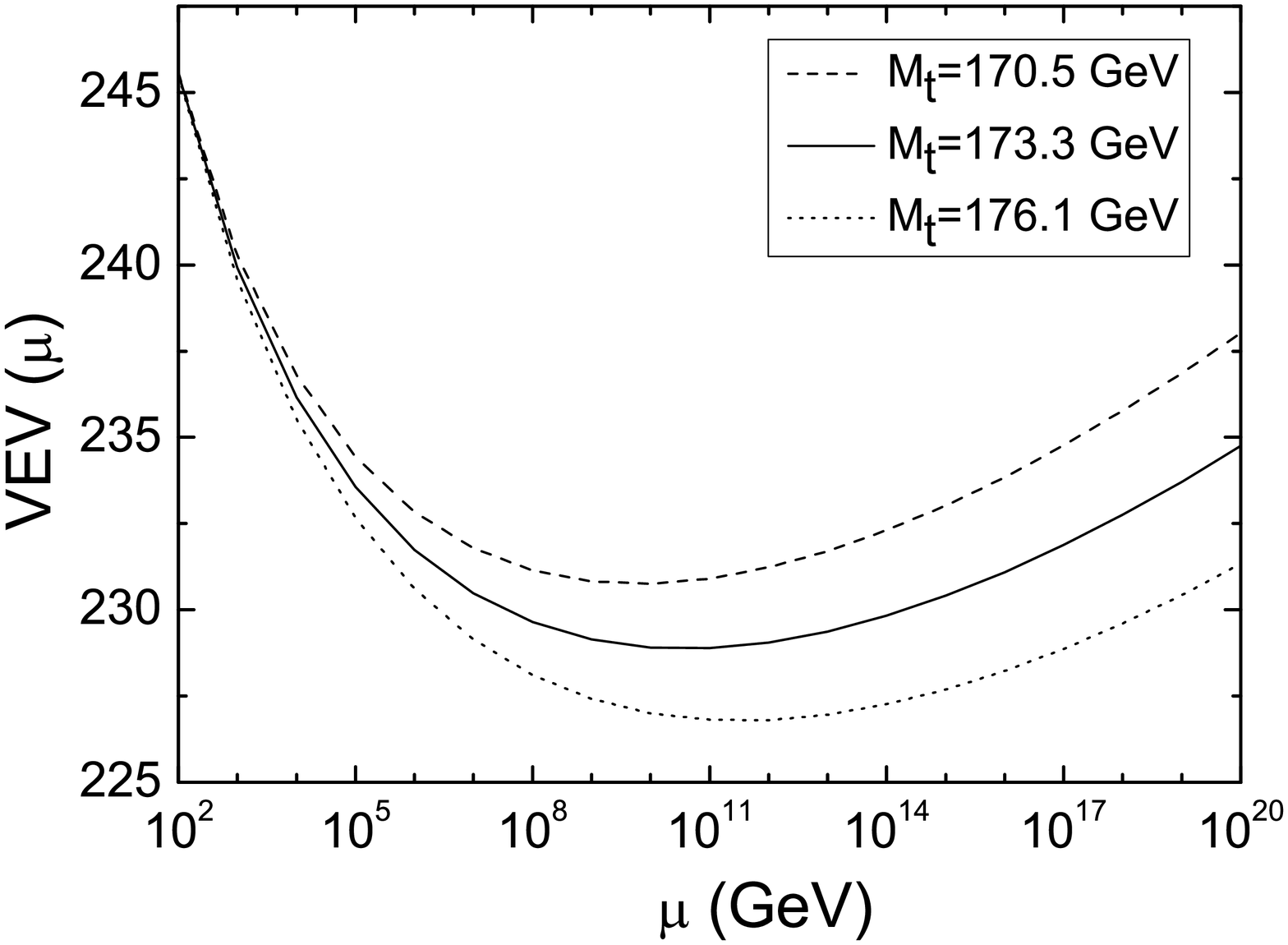}\\
  \caption{The behaviour of the VEV with respect to energy scale up to two-loop level.}\label{fig:vevmu}
\end{figure}
From the Fig.~\ref{fig:vevmu}, we find that the VEV varies very slowly with the energy scale growing, where $\beta$ functions of couplings of the
SM are considered up to two-loop order. When we solve beta functions~\cite{Luo:2002ey,Machacek:1983fi},
the boundary (matching) conditions~\cite{Hambye:1996wb,Hempfling:1994ar,Holthausen:2011aa}
(matching of $\overline{\rm{MS}}$ coupling constants and pole masses
to give boundary conditions of couplings
) are used as in section~\ref{sec:mH2d}.
And the boundary condition of
Eq.~(\ref{eq:advev}) is chosen to be
$v(M_{W})=246.22$ GeV~\cite{Arason:1991ic}
with $M_{W}$ being the pole
mass of the W boson. The behavior of
$\lambda$ with respect to the energy scale $\mu$ is shown in Fig.~\ref{fig:scalarv}.
\begin{figure}
  \centering
  \includegraphics[width=0.6\textwidth]{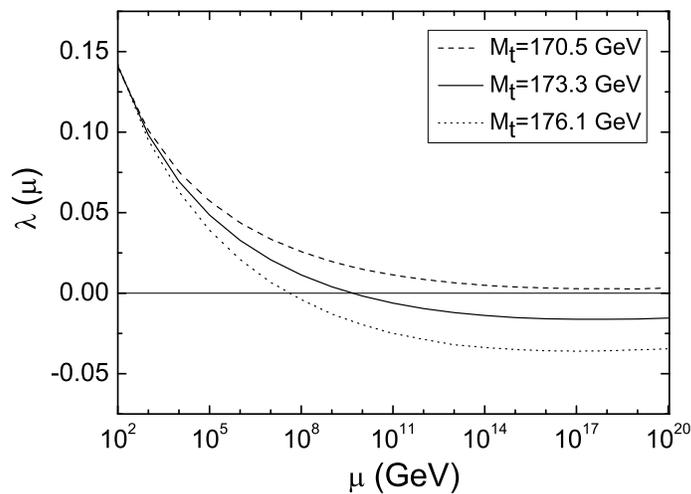}\\
  \caption{The behavior of the scalar quartic coupling with respect to energy
  scale up to two-loop order.}\label{fig:scalarv}
\end{figure}

We present the composed Higgs mass value as a function of energy scale in
Fig.~\ref{fig:mHv} corresponds to different $\lambda$ and $v$.
It can be seen that when we take $M_{t}=173.3$ GeV, the composed Higgs mass
drops gradually to zero at about $10^{10}$ GeV as one expected,
since the behavior of the composed Higgs mass-square
$m^{2}_{H}$ is almost dominated by $\lambda$.\footnote{
The scalar quartic coupling damps gradually as in Fig.~\ref{fig:scalarv}
and the VEV has tiny variations with the energy scale growing, as depicted in Fig.~\ref{fig:vevmu}.}
And the composed Higgs mass-square becomes negative at the energy scale above
10$^{10}$ GeV up to two-loop level which
is related to the vacuum stability argument~\cite{Alekhin:2012py,Lindner:1988ww}.
 \begin{figure}[!htp]
  \centering
  \includegraphics[width=0.7\textwidth]{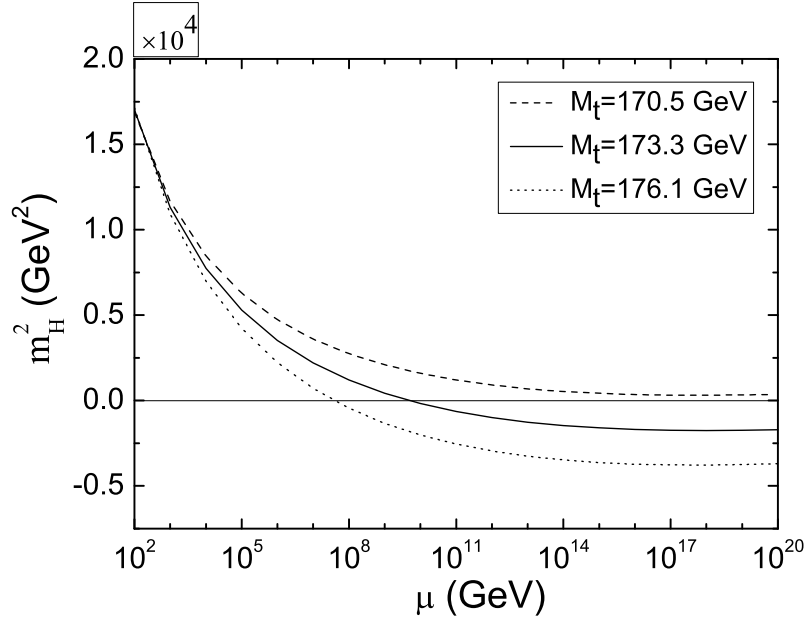}\\
  \caption{The value of the composed Higgs mass-square with energy scale increasing.}\label{fig:mHv}
 \end{figure}
\begin{figure}[!htp]
  \centering
  \includegraphics[width=0.45\textwidth]{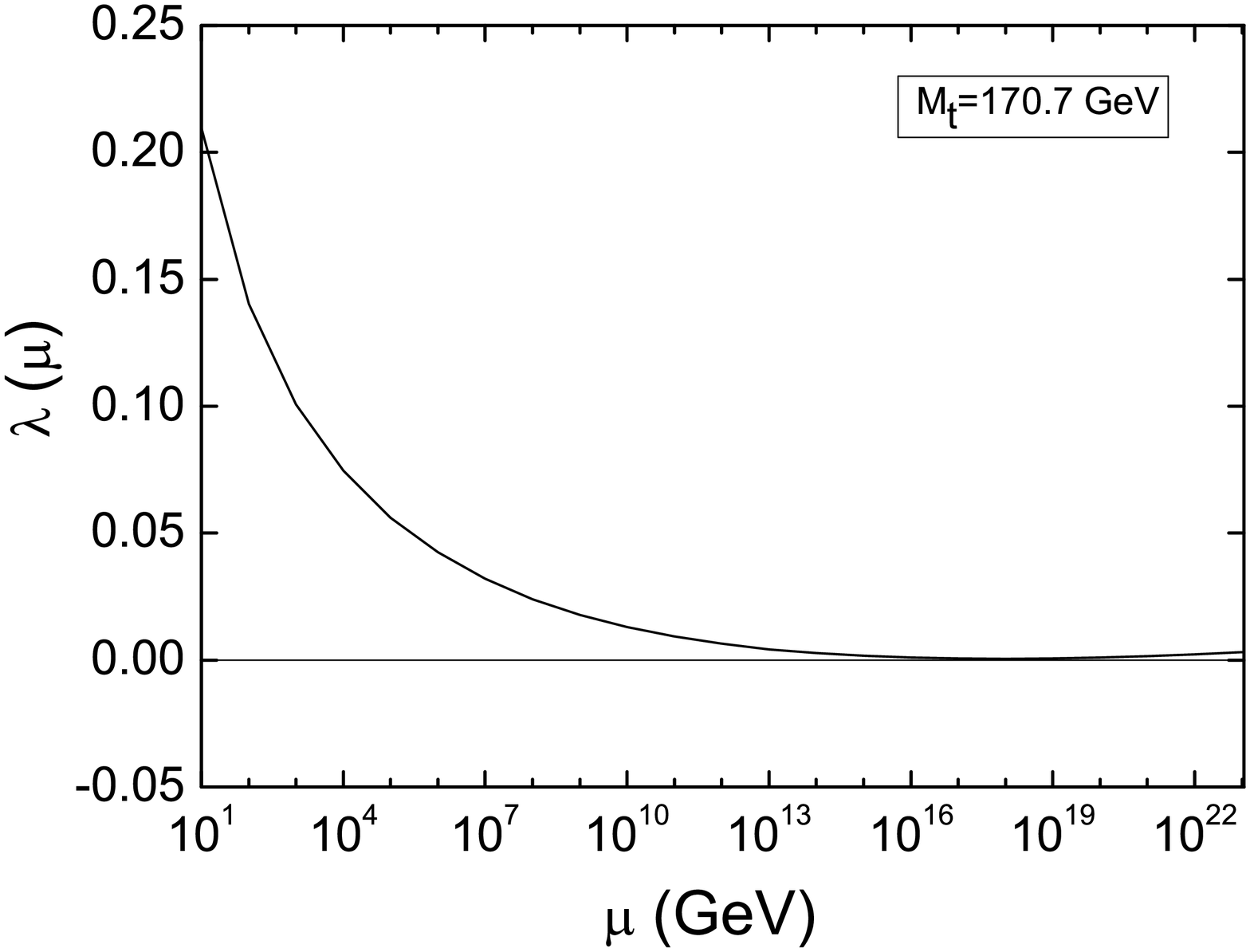}
  \includegraphics[width=0.45\textwidth]{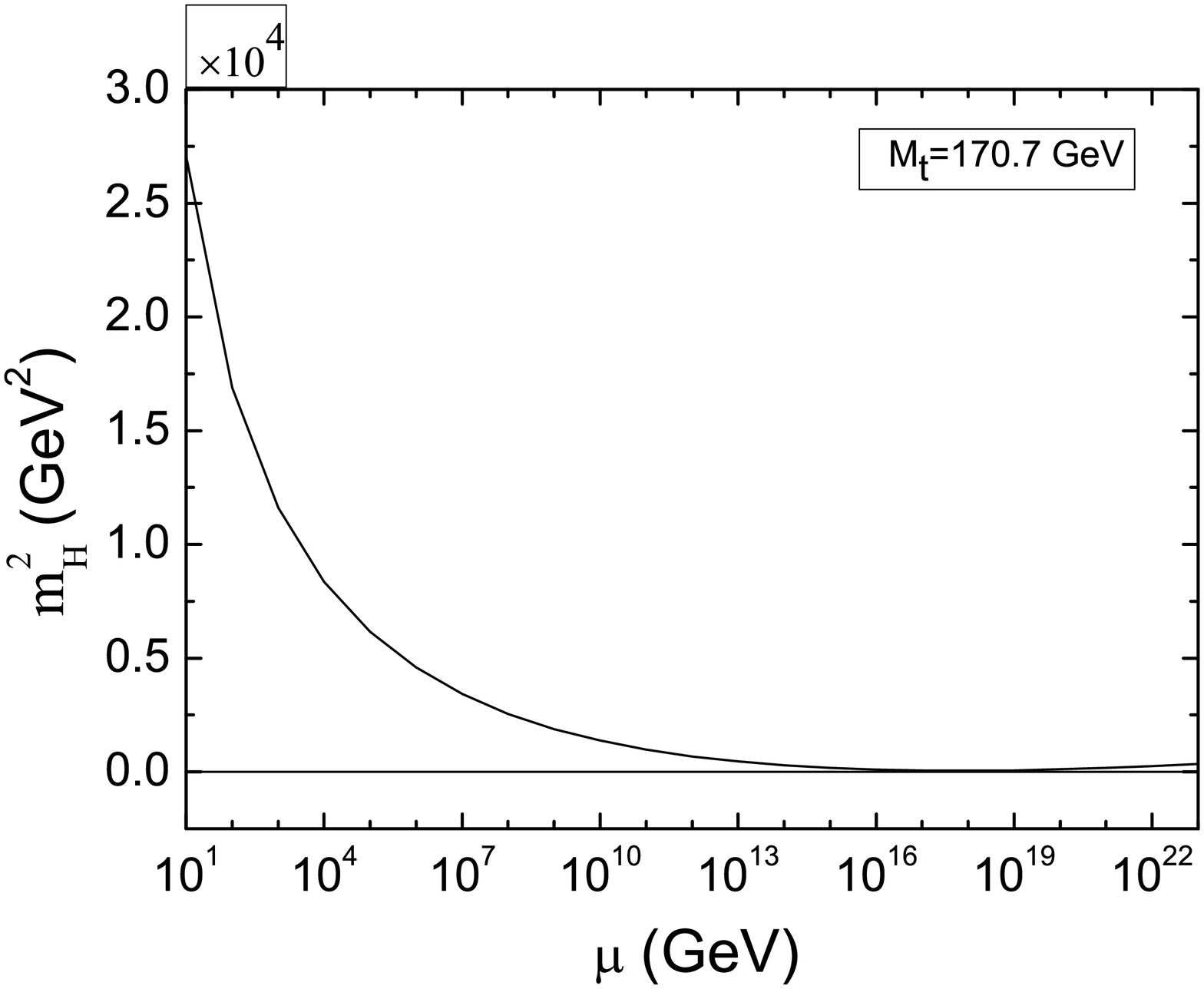}\\
  \caption{Left: Behavior of Higgs quartic coupling with respect to energy scale;
  Right: Behavior of the composed Higgs mass-square with respect to energy scale.}\label{fig:stability}
 \end{figure}
The behaviors of $\lambda, m^2_{H}$ are both strongly dependent of
the top quark pole mass.
The vacuum is always stable when the top quark pole mass is taken to be smaller than about $M_{t}=170.7$ GeV,
and the Fig.~\ref{fig:stability} has shown this critical scenario.
In this case, the $m^2_{H}$ has small positive values at high energy scale.
And, since the line of
$\lambda-\mu$ up to two-loop order is very close
to the line of $\lambda-\mu$ up to three-loop order~\cite{Zoller:2012cv},
the composed Higgs mass-square
$m^{2}_{H}$ up to three-loop order should has almost the same behavior
with the energy scale increasing.

\section{Conclusions}
\label{se:conclusion}

From our analysis, the naturalness problem induced
by quadratic divergences
can live on the complex two dimensional plane,
and the corresponding physics
might does do not affect the scaling property of the RGE
of $m_{H}$ derived at $d=4$ with DREG based on $\rm{MS}(\overline{\rm{MS}})$ scheme.
Suppose one UV complete fundamental theory does exist,
then quadratic divergences in the SM can be compensated by
the same kind of divergences arising from integrations above the
cut-off one used.
As found by us, we can study the quadratic divergences
on the complex two dimensional plane in a gauge invariant scheme,
thus opens a new window to explore UV complete theory.
With more knowledge of UV complete theory, more detailed quadratic divergences
structure of the theory can be studied on the complex two dimensional plane.
Therefore, we can expect to achieve one deeper understanding of the naturalness problem.

The meaning of the Higgs mass term of the SM is revisited based
on the viewpoint of Wilsonian renormalization group. We derived the
RGE of the VEV of the SM in Landau gauge.
Numerical analysis up to two-loop order shows that
the composed Higgs mass damps gradually
with the energy scale growing and eventually to zero
within the valid energy
region of the SM constrained by vacuum stability condition.
And the composed Higgs mass keeps positive even up to the Planck scale
for $M_{t}\leq$ 170.7 GeV.
And in this case,
the gauge invariant property of quantum corrections of the
composed Higgs mass is absent.

\appendix

\section{Integration formulas in divergences calculations }
\label{sec:IDC}

Scalar and tensor integrals involved in one-loop calculations on the complex
two dimensional plane,
 \begin{eqnarray}
\int\frac{d^{d}k}{(2\pi)^{4}}
\frac{k^{\mu}k^{\nu}}{(k^{2}-m^{2})^{2}}
&\rightarrow&-\frac{g^{\mu\nu}}{2}\frac{i}{4\pi}
\frac{1}{1-d/2}\nonumber\\
\int\frac{d^{d}k}{(2\pi)^{d}}
\frac{k^{4}}{(k^{2}-m^{2})^{3}}
&\rightarrow&-\frac{i}{4\pi}\frac{1}{1-d/2}.
\end{eqnarray}

\section{Beta functions of couplings of the SM}
\label{sec:beta}

The beta function for a generic coupling \(x\) is given as:
\begin{align}
\mu\frac{\mathrm{d}x}{\mathrm{d}\mu}&=\beta_x,
\end{align}
The list of beta functions up to two-loop order are given
below~\cite{Einhorn:1992um,Luo:2002ey,Machacek:1983fi}:
\begin{eqnarray}
\beta_{\lambda}&=&\frac{1}{16\pi^2}
\left(\lambda(-9g^2_2-3g^2_1+12g_t^2)+24\lambda^2+\frac{3}{4}g_2^4
+\frac{3}{8}(g_1^2+g_2^2)^2-6g_t^4\right) \nonumber\\
&&+\frac{1}{(16\pi^2)^2}\bigg(-312 \lambda^3 - 144 \lambda^2 g_t^2 + 36 \lambda^2 (3 g_2^2 + g_1^2) -
3 \lambda g_t^4 + \lambda g_t^2 \left(80 g_{3}^2 + \frac{45}{2} g_{2}^2 + \frac{85}{6} g_{1}^2\right)\nonumber\\
&&\quad - \frac{73}{8} \lambda g_2^4 + \frac{39}{4} \lambda g_2^2 g_1^2 + \frac{629}{24} \lambda g_1^4 + 30 g_t^6
 - 32 g_t^4 g_3^2 - \frac{8}{3} g_t^4 g_1^2 - \frac{9}{4} g_t^2 g_2^4 \nonumber\\
&&\quad+ \frac{21}{2} g_t^2 g_2^2 g_1^2 - \frac{19}{4} g_t^2 g_1^4 + \frac{305}{16} g_2^6 - \frac{289}{48}
 g_2^4 g_1^2 - \frac{559}{48} g_2^2 g_1^4 - \frac{379}{48} g_1^6\bigg), \\
\beta_{g_{t}}&=&\frac{1}{16\pi^2}\left(\frac{9}{2}g_t^3+g_t\left(-\frac{17}{12}g_1^2-\frac{9}{4}g_2^2-8g_3^2 \right)\right)\nonumber\\
&&+\frac{1}{(16\pi^2)^2}g_t \bigg(-12 g_t^4 + g_t^2 \left(\frac{131}{16} g_1^2 + \frac{225}{16} g_2^2
+ 36 g_3^2 - 12 \lambda\right) + \frac{1187}{216} g_1^4  \nonumber\\
&&\quad- \frac{3}{4} g_2^2 g_1^2 + \frac{19}{9} g_1^2 g_3^2 -\frac{23}{4} g_2^4 + 9 g_2^2 g_3^2 - 108 g_3^4 + 6 \lambda^2\bigg), \\
\beta_{g_{1}}&=&\frac{1}{16\pi^2}\big(\frac{41}{6}g_1^3\big)
+\frac{1}{(16\pi^2)^2}g_1^3 \left(\frac{199}{18} g_1^2 + \frac{9}{2} g_2^2 + \frac{44}{3} g_3^2 - \frac{17}{6}g_t^2\right),\\
\beta_{g_{2}}&=&-\frac{1}{16\pi^2}(\frac{19}{6}g^3_2)
+\frac{1}{(16\pi^2)^2}g_2^3 \left(\frac{3}{2} g_1^2 + \frac{35}{6} g_2^2 + 12 g_3^2 -\frac{3}{2}g_t^2\right),\\
\beta_{g_{3}}&=&\frac{1}{16\pi^2}\big(-7g_3^3\big)
+\frac{1}{(16\pi^2)^2}g_3^3
\left(\frac{11}{6} g_1^2 + \frac{9}{2} g_2^2 - 26 g_3^2 -2 g_t^2\right).
\end{eqnarray}

\end{document}